# Investigation of a hydraulic impact

# – a technology in rock breaking

**Martin Genet**[*]

*Laboratoire de Mécanique et de Technologie,*

*61 av. du Président Wilson, 94235 Cachan Cedex, France*

**Dr. Wenyi Yan**

*Senior Lecturer*

*Department of Mechanical Engineering*

*Building 31, Monash University, VIC800, Australia*

**Pr. Thanh Tran-Cong**

*RME Chair in Computational Engineering,*

*Computational Engineering and Science Research Centre,*

*University of Southern Queensland, Toowoomba, Australia*

*Corresponding author. Tel: 0033147402225, E-mail: martin.genet@lmt.ens-cachan.fr

# Abstract

The finite element method and dimensional analysis have been applied in the present paper to study a hydraulic impact, which is utilized in a non-explosive rock breaking technology in mining industry. The impact process of a high speed piston on liquid water, previously introduced in a borehole drilled in rock, is numerically simulated. The research is focused on the influences of all the parameters involved in the technology on the largest principal stress in the rock, which is considered as one of the key factors to break the rock. Our detailed parametric investigation reveals that the variation of the isotropic rock material properties, especially its density, has no significant influence on the largest principal stress. The influences of the depth of the hole and the depth of the water column are also very small. On the other hand, increasing the initial kinetic energy of the piston can dramatically increase the largest principal stress and the best way to increase the initial kinetic energy of the piston is to increase its initial velocity. Results from the current dimensional analysis can be applied to optimize this non-explosive rock breaking technology.

**Key words** Finite element simulation, Dimensional analysis, Rock breaking, Non-explosive method, Hydraulic impact.

# 1 Introduction

Explosives are commonly used to fragment large rock masses in modern mining practice. From the technical point of view, although explosive method is powerful, it does not produce fragments with homogeneous size distribution. In many situations the amount of very fine rocks is high, while in other situations the amount of oversized boulders could be excessive. Furthermore, explosive method involves complex drilling, blasting, scaling, ground support and the evacuation of people and equipment before blasting. Such a multi-activity cycle is time-consuming, inefficient and unproductively expensive [1]. Another major concern with explosive blasting is the associated danger and undesirable impact on the environment such as fly rocks, air blast, noise pollution and toxic fumes. When blasting occurs close to residential areas, or during tunnel construction, environmental protection regulation could seriously affect the rate of rock excavation. In some cases, blasting would be excluded as an acceptable method of rock breaking. Apart from the breaking of large rock masses for transportability purposes, tunnelling requires more carefully controlled rock breaking. Oversized boulders often cause blockage of mine draw points. When such blockage occurs, extensive shutdown of mine operation will result, causing loss of millions of dollars per hour. Thus, fast, simple, safe and clean methods of breaking boulders are required in some cases to make total mining operation efficient.

To overcome the drawbacks of explosive methods, several non-explosive technologies have been developed in the past, see Singh [1]. Young [2] provided an overview and compared the pros and cons of various methods of pressurising a borehole in a rock mass, including small charge explosive and propellant, water jets, firing of high speed water slugs, mechanical splitters, and high pressure gases. McCarthy [3] proposed the use of propellant cartridge in a predrilled hole to provide the breaking force. In the latest patent by Young [4], a high-pressure foam was utilized to replace explosive technology. The controlled-foam injection (CFI) method invented by Young [1, 4] has a number of advantages, including lower maximum pressure and the maintenance of pressure during fracture by virtue of the compressibility of the foam. Another non-explosive rock breaking method was invented and patented by Denisart et al. [5], which is illustrated in Fig. 1. The hydraulic fluid, such as water, is introduced in a pre-drilled borehole and is impulsively loaded by a high speed piston. The highly pressurized water, with the reflection of the pressure wave, will result in huge stresses in the rock mass, especially at the bottom of the hole, which is a stress raiser. Consequently, cracks will be initiated around the borehole, especially at the bottom, and the pressurized water will penetrate into the cracks, providing the driving force for crack propagation. Eventually, cracks will propagate back to the surface due to free surface effect and a volume of the hard material will be removed.

All the available inventions and patents focused on the principle of the rock breaking methods, i.e., different approaches are used to answer the basic question of how to break a rock mass. In terms of the application of the non-explosive technology, we need to quantify many parameters, such as the depth of the borehole and the initial velocity of the piston, etc. In the current investigation, the dimensional analysis and numerical method are applied to quantify the hydraulic impact process, which is involved in a non-explosive rock breaking technology as shown in Fig. 1. The efficiency of the impact is evaluated by the maximum principal stress in the rock during an impact process. Numerical results from the investigation will assist industry to quantitatively apply this non-explosive rock breaking technology and, therefore, to improve rock breaking efficiency.

This paper is structured as follows. The finite element model to simulate the hydraulic impact is presented in Section 2. The functional relationship between the largest principal stress in the rock and all the processing parameters from dimensional analysis is discussed in Section 3. In Section 4, detailed numerical results on the influences of all the parameters on the largest principal stress are discussed. Finally, conclusions are given in Section 5.

## 2 Impact simulation

### 2.1 Finite element mesh

The finite element method has been applied to simulate the hydraulic impact process. We use a finite element package, CASTEM, to create the finite element mesh, a PERL script to translate CASTEM meshes into ABAQUS meshes, the commercial package ABAQUS/Explicit to do the simulations, and a PYTHON script to automate the simulation process.

The borehole, as shown in Fig. 1, can be idealized as a cylinder. Therefore, this problem can be treated as axisymmetric. Moreover, the rock body can be considered as semi-infinite. Infinite elements are utilized to simulate the semi-infinite body. Fig. 2 shows the finite element mesh generated by using CASTEM. The definition of the geometrical parameters shown in Fig.2 can also be found in Table 1.

As shown in Fig. 2, there is an arc with the radius of $R_a$ at the bottom of the borehole. The reason to introduce the arc is to avoid singularity, which implies an infinite stress. Moreover, it is a good representation of the reality. It is impossible to have a perfect right angle practically when we drill a hole and we always have the trace of the tool on the machined part. Further explanation of this assumption can be found in Section 3.

As shown in Fig. 2, very fine meshes are generated around the corner of the bottom of the hole. Many simulation tests with different mesh densities have been carried out to eliminate the influence of mesh density and determine the final mesh for the calculations. In the end, there are 122 four-node bilinear elements and 4349 three-node linear elements in the final model. The only initial condition in this problem is the initial velocity of the piston, which is an additional parameter of the problem.

## 2.2 Material properties

The piston is normally made of steel, which is assumed to be a homogeneous and isotropic material. We also assume that the piston stays in its elastic domain during the impact process. Thus we can choose the piston's material data as Young's modulus $E_p = 200\ GPa$, Poisson's ratio $v_p = 0.3$ and density $\rho_p = 7800\ Kg/m^3$. During the transient impact process, the water in the borehole can be considered as still, i.e., no flow. According to Wilson [6], we can model the water as an elastic, homogeneous and isotropic solid with theses material data: Young's modulus $E_w = 6207813\ Pa$, Poisson's ratio $v_w = 0.4995$ and density $\rho_w = 1000\ Kg/m^3$.

In the current investigation, the rock in this simulation is simplified as an elastic, homogeneous and isotropic solid. In reality, rock, as a natural material, consists of crystal, grains, cementitious materials, voids, pores and flaws, see [7]. At the first stage of investigating this non-explosive rock breaking technology, our current objective is to understand and quantify the impact process. Considering the influence of the inhomogeneous microstructure of rock material will be our next task. On the other hand, because of the uncertainty of the microstructure and its inhomogeneity, the assumed isotropic rock in our model can be treated as a representative of the real material and the results from this assumption will still be practically useful, especially for companies which intend to develop relevant universe equipments for this non-explosive technology. Furthermore, we do not take into account the possible plastic deformation of the rock in the present research, neither the creation nor the propagation of cracks, which will be our future study. Consequently, we have three parameters to describe the rock: Young's modulus $E_r$, Poisson's ratio $v_r$ and it's density $\rho_r$.

## 2.3 Contact simulation

As shown in Fig. 1 and Fig. 2(a), there are three pairs of contacts involved in the impact process, i.e., the contact between the piston and the water, the contact between the piston and the rock, and the contact between the water and the rock. We use the hard contact algorithm from ABAQUS without damping to simulate these contacts. The friction is also neglected in our simulation. Practically, the friction between water and piston or rock should be very low. Further study will be carried out after we obtain reliable friction value involved in the contact between piston and rock.

All the impact simulations were carried out by using ABAQUS/Explicit. The effect of the hydraulic impact is evaluated by the largest principle stress in the rock. As an example of our finite element simulation results, Fig. 3 shows the distribution of the maximum principal stress field in the structure at the instant when the shockwave arrives at the end of the bottom of the hole. It clearly indicates that the largest maximum principal stress, the largest principal stress in short, occurs at the bottom of the hole. Irrespective of microstructures, cracks will possibly initiate at this position with this largest stress in the rock. Fig. 4 shows the corresponding direction field of the maximum principal stress at this local area in the rock. On the surface, the direction of the maximum principal stress is perpendicular to the surface. One can imagine, once cracks initiate, the highly pressurized water will penetrate into the cracks and drive the cracks to propagate, which will be the core of the investigation to understand the rock breaking in our future study.

# 3. Dimensional analysis

All the parameters involved in the simulation are listed in Table 1. Here, the radius of the piston is the same as the radius of the borehole. Dimensional analysis is a powerful method to systematically carry out parametrical study on a complicated problem involving many parameters, see examples [8-10]. This method is applied in the current investigation. The objective variable in our dimensional analysis is chosen as the largest principal stress in the rock, $\sigma_m$, during an impact process. The rock material can be roughly considered as brittle material. According to Coulomb's criterion of maximum normal stress, the largest principal stress will initiate cracks in the rock and lead to the fragmentation of rock mass. Generally, the largest principal stress is a function of all the parameters listed in Table 1, i.e.,

$$\sigma_m = f\left(L_p, E_p, \nu_p, \rho_p, V_p, D_w, E_w, \nu_w, \rho_w, R_h, D_h, R_a, E_r, \nu_r, \rho_r\right). \tag{1}$$

According to the Buckingham $\Pi$-theorem for dimensional analysis, we can reduce the number of parameters. For this purpose, we choose $D_h$, the depth of the hole, $\rho_p$, the density of the piston and $E_p$, the Young modulus of the piston as the primary quantities. Therefore, the dimensionless function for the largest principal stress is

$$\frac{\sigma_m}{E_p} = \Pi_1\left(\frac{L_p}{D_h}, \nu_p, \frac{V_p}{E_p^{1/2} \times \rho_p^{-1/2}}, \frac{D_w}{D_h}, \frac{E_w}{E_p}, \nu_w, \frac{\rho_w}{\rho_p}, \frac{R_h}{D_h}, \frac{R_a}{D_h}, \frac{E_r}{E_p}, \nu_r, \frac{\rho_r}{\rho_p}\right). \tag{2}$$

Among all the dimensionless parameters that we have just created, some values can be considered as unchanged in this physical problem. The piston is generally made from steel and water is normally used as the liquid in this technology. Therefore, the material data for the piston and the water can be treated as constant. Consequently, the following dimensionless parameters will be considered constant in our model:

$$\nu_p = 0.3, \qquad\qquad \nu_w = 0.4995, \tag{3}$$

$$\frac{E_w}{E_p} = \frac{6207813}{2\times 10^{11}} = 31\times 10^{6}, \qquad \frac{\rho_w}{\rho_p} = \frac{1000}{7800} = 0.128. \qquad (4)$$

Therefore, the dimensionless function (2) can be simplified as

$$\frac{\sigma_m}{E_p} = \Pi_2 \left( \frac{L_p}{D_h}, \frac{V_p}{E_p^{1/2} \times \rho_p^{-1/2}}, \frac{D_w}{D_h}, \frac{R_h}{D_h}, \frac{R_a}{D_h}, \frac{E_r}{E_p}, \nu_r, \frac{\rho_r}{\rho_p} \right). \qquad (5)$$

After this dimensional analysis, the number of variables involved in the stress analysis has reduced from 15 in the original Eq. (1) to 8 in Eq. (5).

We now define the domains of the dimensionless variables based on our understanding of this physical problem. The following limits of the domains for geometrical parameters are appropriated for this problem:

$$\frac{L_p}{D_h} \in [0.1; 0.8], \qquad \frac{D_w}{D_h} \in [0.1; 0.8]. \qquad (6)$$

$$\frac{R_h}{D_h} \in [0.01; 0.5], \qquad \frac{R_a}{D_h} \in [0.001; 0.05]. \qquad (7)$$

Referring to [11], the domains of the mechanical properties of different types of rocks are shown in Table 2. According to this table, after choosing $E_p = 200 \; GPa$ and $\rho_p = 7800 \; Kg/m^3$, we can define the domains for dimensionless variables linked to the rock material as follows:

$$\frac{E_r}{E_p} \in [0.05; 0.5], \qquad \nu_r \in [0.1; 0.35], \qquad \frac{\rho_r}{\rho_p} \in [0.25; 0.4]. \qquad (8)$$

According to Denisart et al. [5], the initial velocity of the piston can vary from $10 \; m/s$ to $200 \; m/s$, which corresponds to the following domain of the dimensionless initial velocity:

$$\frac{V_p}{E_p^{1/2} \times \rho_p^{-1/2}} \in [0.0020; 0.0395]. \qquad (9)$$

In our numerical simulations, we have to adjust the limits of some parameters' domains because of numerical instability problem. The Poisson's ratio of water is 0.4995, which is close to the value of 0.5 of imcompressible materials. Additionally, the water is highly confined in the borehole and exposed to a highly compressive load from the impact of the high speed piston. Due to these factors, the stiffness matrix in a finite element simulation is almost singular, which can sometimes lead to numerical instability [12]. A unsuccessful numerical instability calulcation can be easily detected by observing large abnormally distorted and penetrated deform meshes. To overcome this instablility problem, we sometimes have to choose some values lower than the upper limit or higher than the lower limit of the domains defined in above Eqs (6-9). In the following section, only the correct results from the stable calculations are reported. We can believe that the fitted laws in the restricted domains of study in the following section are valid in the entire domains defined in Eqs (6-9).

## 4. Results and discussions

### 4.1. Influence of rock properties

The effect of rock density is considered first. Fig. 5 shows the influence of the normalized rock density, $\rho_r/\rho_p$, on the normalized largest principal stress, $\sigma_m/E_p$, in the rock during the impact process. We have considered the two limit values of the domains of all the dimensionless parameters in Eq. (5), one by one from Fig. 5(a) to Fig. 5(g), while fixing the values of all the others dimensionless parameters at the middle values of their domains, which are defined in Eqs. (6-9). For example, the two curves in Fig. 5(a) are obtained for $L_p/D_h = 0.17$ and $L_p/D_h = 0.58$ respectively while fixing $v_r = 0.19$, $E_r/E_P = 0.3$, $R_a/D_h = 0.026$, $R_h/D_h = 0.26$, $D_w/D_h = 0.5$ and $V_p/(\rho_p^{-1/2}E_p^{1/2}) = 0.013$.

Figs. 5(a-g) clearly indicate that the variation of the normalized largest principal stress $\sigma_m/E_p$ due to the change of the normalized rock density $\rho_r/\rho_p$ from 0.25 to 0.4 in all the studied cases is negligibly small. Because all the studied cases have covered the domains of this physical problem, one can deduce that it is generally correct that the influence of the variation of rock density on the maximum largest principal stress in the rock can be neglected. Consequently, the dimensionless rock density in Eq. (5) can be removed and the dimensionless stress function can be further simplified as

$$\frac{\sigma_m}{E_p} = \Pi_3 \left( \frac{L_p}{D_h}, \frac{V_p}{E_p^{1/2} \times \rho_p^{-1/2}}, \frac{D_w}{D_h}, \frac{R_h}{D_h}, \frac{R_a}{D_h}, \frac{E_r}{E_p}, \nu_r \right). \tag{10}$$

Fig. 6 shows the numerical results of the normalized largest principal stress in the rock during the impact for different values of the Poisson's ratio of the rock. Similarly, all the others dimensionless parameters in Eq. (10) are fixed at their middle values of their domains, and only the Poisson's ratio of the rock changes from one calculation to another. We can see that the normalized largest principal stress increases slightly with the increasing of the Poisson's ratio of the rock. For example, if $E_p = 200\ GPa$, $\rho_r = 7800\ Kg/m^3$, $D_h = 60\ cm$, $L_p = 20\ cm$, $V_p = 60\ m/s$, $D_w = 30\ cm$, $R_h = 5\ cm$, $R_a = 5\ mm$ and $E_r = 50\ GPa$, according to Fig. 6, then the largest principal stress in the rock increases from 862 $MPa$ to 954 $MPa$ when the Poisson's ratio of the rock evolves from 0.1 to 0.35. As shown in Fig. 6, the set of the numerical data can be well fitted by the following exponential function:

$$\frac{\sigma_m}{E_p} = 0.0043 + 0.0107 \times \nu_r^{2.98}. \tag{11}$$

Fig. 7 shows the variation of the normalized largest principal stress in the rock during the impact for different normalized values of the Young's modulus of the rock. Similar to Fig. 6, $\sigma_m / E_p$ increases slightly with the increasing of $E_r / E_p$ from 0.1 to 0.5. For example, if $E_p = 200\ GPa$, $\rho_r = 7800\ Kg/m^3$, $D_h = 60\ cm$, $L_p = 20\ cm$, $V_p = 60\ m/s$, $D_w = 30\ cm$, $R_h = 5\ cm$, $R_a = 5\ mm$ and $\nu_r = 0.2$, we see from Fig. 7 that the largest principal stress in the rock evolves from $746\ MPa$ to $934\ MPa$ when the Young's modulus of the rock evolves from $10\ GPa$ to $100\ GPa$. The set of the numerical data is also fitted by an exponential function, which is shown with the thick curve in Fig. 7.

## 4.2. Influence of borehole dimensions

### 4.2.1 Borehole depth

The depth of the borehole, $D_h$, is chosen as the primary length in the dimensional analysis. Its influence on the problem can be implicitly reflected in the parametric study of other dimensionless length parameters, such as the dimensionless piston length and the dimensionless water depth. But it is understandable that the depth of the borehole has no direct influence on the largest principal stress in the rock, and that is the reason to choose it as the primary length to normalize the other parameters.

### 4.2.2 Borehole radius

Fig. 8 shows the influence of the normalized borehole radius $R_h / D_p$ on the normalized largest principal stress in the rock $\sigma_m / E_p$ while other parameters are fixed at the middle values of their domains. It indicates that $\sigma_m / E_p$ increases gradually with $R_h / D_p$. Bear in mind, the borehole radius is equal to the radius of the water column and the radius of the piston. Increasing borehole radius means increasing the radius of the piston, and therefore, increasing the initial kinetic energy of the piston with fixed initial velocity. This influence in real value is very significant. For example, if $E_p = 200\ GPa$, $\rho_r = 7800\ Kg/m^3$,

$D_h = 60\ cm$, $L_p = 20\ cm$, $V_p = 60\ m/s$, $D_w = 30\ cm$, $E_r = 50\ GPa$, $R_a = 5\ mm$ and $v_r = 0.2$, then the largest principal stress in the rock evolves from $842\ MPa$ to $2244\ MPa$ when the borehole radius evolves from $5\ cm$ to $30\ cm$ according to the numerical results.

It is interesting to plot the instantaneous average velocity and the instantaneous kinetic energy of the piston over the impacting time for several values of the dimensionless radius of the hole. Fig. 9 and Fig. 10 show these results. In Fig. 9, we can see that at the beginning, the piston is moving down, that is why the average velocity is negative. Then its average velocity decreases, and at $t / D_h \times \rho_p^{1/2} \times E_p^{-1/2} = 7$, the piston has zero averaged velocity. After that, the piston is coming up, so its average velocity is increasing. We can also follow this in Fig. 10: the piston starts with its kinetic energy, which decreases until $t / D_h \times \rho_p^{1/2} \times E_p^{-1/2} = 7$ where it is null, and then increases because the piston is coming up. Additionally, we can see from Fig. 9 that the average velocity of the piston does not depend on the borehole radius, and from Fig. 10 that the kinetic energy of the piston depends on the radius of the hole, which is obvious because the radius of the hole is also the radius of the piston and the initial kinetic energy strongly depends on the radius of the piston. However, we can see that its evolution is quite similar from one value to another, which is obviously linked to the fact that the average velocity is the same for all curves in Fig. 9. We can explain this by saying that both the kinetic energy of the piston and the energy that is transmitted to the water depend on the radius of the piston in the same way: they both are proportional to the cross section area of the piston, i.e., the square of its radius. Then, if the piston has a larger radius, it will have more initial kinetic energy, but it will also transmit more energy to the water, so its kinetic energy will decrease faster. This remark is important, and we will explain below that the kinetic energy of the piston is a key factor to determine the largest principal stress in the rock.

To understand this, we discuss an impact process by following Fig. 11, which shows the first principal stress in the element that has the largest principal stress and the instantaneous average velocity of the piston versus impacting time for $R_h/D_h = 0.0167$:

- The piston hits the water with its initial velocity at the beginning of the simulation, which starts the impact process.

- The created shockwave comes down in the water, and arrives at the bottom of the borehole at $t/(D_h \times \rho_p^{1/2} \times E_p^{-1/2}) = 2$. Here is the first peak in the stress curve, with $\sigma/E_p = 1 \times 10^{-3}$.

- Then the shockwave climb back to the surface (the material behaviour of the water is very different from the others material behaviours, so the transmission of energy is low) and the stress at the bottom of the hole reduces because of the dispersion of energy.

- When the shockwave arrives at the surface at $t/(D_h \times \rho_p^{1/2} \times E_p^{-1/2}) = 4$, it is reflected and comes down again, but with more energy because the piston is still coming down.

- Thus, the second peak in the stress curve will be greater, with $\sigma/E_p = 1.7 \times 10^{-3}$ at $t/(D_h \times \rho_p^{1/2} \times E_p^{-1/2}) = 6$.

- And when the shockwave climb back and arrives at the surface for the second time, at $t/(D_h \times \rho_p^{1/2} \times E_p^{-1/2}) = 8$, the piston has no more velocity downward and is going up.

- So the shockwave comes down again with less energy, and the third peak in the stress curve will be smaller, with $\sigma/E_p = 1.6 \times 10^{-3}$ at $t/(D_h \times \rho_p^{1/2} \times E_p^{-1/2}) = 10$.

We can conclude that we have the largest principal stress in the rock when the shockwave arrives at the bottom of the hole the last time while the piston still having some velocity downward. It indicates that the kinetic energy of the piston plays an important role in the determination of the largest principal stress.

**4.2.3 Arc radius at borehole bottom**

This parameter, the arc radius at the borehole bottom, as shown in Fig. 2, is introduced to avoid the problem of infinite stress at the corner and allow a better modelling of the geometry of the hole in a real situation. Obviously, it will have a substantial influence on the largest principal stress in the rock. Fig. 12 shows its influence on the problem. The thick curve in Fig. 12 represents an exponential function to fit the numerical dots. We can see that the stress tends to infinity when the radius of the arc closes to zero, which is normal because of the problem of singularity when $R_a = 0 \, mm$. Practically, this parameter will never be equal to zero, and its value can be estimated from the documentation of the active part of the tool used to drill the hole, in addition to the consideration of the rock material, or from experimental tests. We can achieve the dimensional values from Fig. 12. For instance, if $E_p = 200 \, GPa$, $\rho_r = 7800 \, Kg/m^3$ and $D_h = 60 \, cm$, $L_p = 20 \, cm$, $V_p = 60 \, m/s$, $D_w = 30 \, cm$, $E_r = 50 \, GPa$, $R_h = 5 \, cm$ and $\nu_r = 0.2$, the largest principal stress in the rock reduces from $1860 \, MPa$ to $538 \, MPa$ when the radius of the arc at the bottom of the hole increases from $1 \, mm$ to $10 \, mm$.

**4.2.4. Water depth**

Fig. 13 shows the influence of $D_w$ on the largest principal stress in the rock and on the time to reach this value, when all the other dimensionless parameters are fixed at the middle values of their domains. We can see that the largest principal stress decreases and that the time to reach that stress increases when the depth of water increases. This conclusion can be explained by the fact that the deeper the water is, the more the energy can disperse from the water to the rock, and therefore, it results to a smaller largest principal stress in the rock at the bottom of the hole. This relationship depends on the distance travelled by the shockwave, and Fig. 13 indicates that the normalized stress curve consists of two linear parts with the corner at $D_w/D_h = 0.35$. Furthermore, the slope of the normalized stress curve depends on the number of return trips that the shockwave has made before the largest principal stress is reached and the change of the curve's slope is linked to the discontinuity of the curve for the time to reach the largest stress, and is explained in detail below.

The variation of the time to reach the maximum principal stress with the change of the water depth, shown in Fig. 13, is due to the combination of two facts:

- when the depth of water increases, the time for the shockwave to travel from the surface of the water to the bottom of the hole increases too, so the time to reach the largest principal stress increases.

- and when the water becomes deep enough, the largest principal stress is not reached at the third time when the shockwave arrives at the bottom of the hole, but the second time. We can follow this in Figs. 14(a) and 14(b), which give the first principal stress in the element that has the largest principal stress in the rock and the kinetic energy of the piston versus times, respectively for $D_w/D_h = 0.333$ and $D_w/D_h = 0.417$, which are respectively around the corner of the stress curve and the dropping part of the time curve in Fig. 13. Fig. 14(a) shows that the largest principal stress is reached at the third time when the shockwave arrives at the bottom of the hole for $D_w/D_h = 0.333$ and Fig. 14(b) shows it is reached at the second time for $D_w/D_h = 0.417$. This is the reason for the time dropping between $D_w/D_h = 0.333$ and $D_w/D_h = 0.417$ and the appearance of the corner of the stress curve in Fig. 13.

Fig. 13 indicates that reducing the water depth can increase the largest stress in the rock. For example, in the case of $E_p = 200\ GPa$, $\rho_r = 7800\ Kg/m^3$, $D_h = 60\ cm$, $L_p = 20\ cm$, $V_p = 60\ m/s$, $R_a = 5\ mm$, $E_r = 50\ GPa$, $R_h = 5\ cm$, $\nu_r = 0.2$, the largest principal stress in the rock increases from $773\ MPa$ to $1207\ MPa$ when the depth of water in the hole reduces from 50 cm to 10 cm. In terms of the entire rock breaking technology, we need pressurized water to drive crack propagation once cracks are initiated in the rock. Therefore, it is not recommended to increase the largest principal stress in the rock, the crack initiation forcing, by reducing the water depth.

### 4.3. Influence of piston dimensions

#### 4.3.1. Piston length

The radius of the piston is the same as $R_h$, the radius of the hole. Its influence has been studied in the previous subsection. Now, let's consider the influence of piston length. Fig. 15 shows that the normalized largest principal stress increases continuously with the increasing of the normalized piston length. The longer the piston is, the higher its initial energy is because of the fixed initial velocity. Therefore, the stress in the rock increases. Such an influence is significant. For example, in the case of $E_p = 200\ GPa$, $\rho_r = 7800\ Kg/m^3$, $D_h = 60\ cm$, $D_w = 30\ cm$, $V_p = 60\ m/s$, $R_a = 5\ mm$, $E_r = 50\ GPa$, $R_h = 5\ cm$ and $\nu_r = 0.2$, we obtain from Fig. 15 that the largest principal stress in the rock evolves from $690\ MPa$ to $1202\ MPa$ if the piston length evolves from $10\ cm$ to $50\ cm$.

Fig. 15 also shows that the time to reach the largest stress has an irregular relationship with the piston length. This can be explained by Figs. 16(a-c), which show the relationships between the normalized first principal stress in the element that has the largest value in the rock during the impact, the normalized kinetic energy of the piston and the normalized impacting time for different values of the normalized piston length, $L_p/D_h = 0.1667$ in Fig. 16(a), $L_p/D_h = 0.4167$ in Fig. 16(b), and $L_p/D_h = 0.5833$ in Fig. 16(c), while fixing the other dimensionless parameters at the middle values of their domains. Fig. 16(a) indicates that when the shockwave climbs back to the surface at the second time when the slope of the curve of the kinetic energy changes, at $t/{D_h \times \rho_p^{1/2} \times E_p^{-1/2}} = 7.5$, the piston has no more velocity downward and is coming up (its kinetic energy has already been null), which means that all its energy has already been transferred to the structure, so the largest principal stress has already been reached, the second time when the shockwave arrives at the bottom of the hole ($\sigma_m/E_p = 3.5 \times 10^{-3}$ at $t/{D_h \times \rho_p^{1/2} \times E_p^{-1/2}} = 6$). It is exactly the same process as in previous Subsection 4.2.2. In the Fig. 16(b), the impacting process is similar: the largest principal stress is also reached at the second time when the shockwave arrives at the bottom of the hole, at $t/{D_h \times \rho_p^{1/2} \times E_p^{-1/2}} = 6$, but due to the higher impacting energy, the value of the normalized largest stress is larger, with $\sigma_m/E_p = 4.5 \times 10^{-3}$. And Fig. 16(c) is for an even longer piston: we see here that the piston still have some downward velocity (its kinetic energy has not been null) when the shockwave returns at the surface for the second time at $t/{D_h \times \rho_p^{1/2} \times E_p^{-1/2}} = 7.5$. Therefore, the largest principal stress is reached at the third time when the shockwave arrives at the bottom of the hole: $\sigma_m/E_p = 5.3 \times 10^{-3}$ at $t/D_h \times \rho_p^{1/2} \times E_p^{-1/2} = 10$. This is the reason why the time to reach the largest principal stress is irregular as shown in Fig. 15.

### 4.3.2. Piston's initial velocity

Fig. 17 shows the variation of the normalized principal stress in the rock with the change of the normalized initial velocity of the piston. It clearly indicates that the normalized stress increases linearly with the normalized velocity and the variation rate is significant. For example, in the case of $E_p = 200\ GPa$, $\rho_r = 7800\ Kg/m^3$, $D_h = 60\ cm$, $D_w = 30\ cm$, $L_p = 20\ cm$, $R_a = 5\ mm$, $E_r = 50\ GPa$, $R_h = 5\ cm$, and $\nu_r = 0.2$, the largest principal stress in the rock evolves from $700\ MPa$ to $2800\ MPa$ when the initial velocity of the piston evolves from $50\ m/s$ to $200\ m/s$.

### 4.3.3. Initial kinetic energy of the piston

We have investigated the influence of the dimensions of the piston and its initial velocity on the largest principal stress in the rock. We will now try to understand the global influence of its initial kinetic energy, which embrace all these parameters:

$$K_p = \frac{1}{2} \times \rho_p \times \pi \times R_p^2 \times L_p \times V_p^2. \tag{12}$$

Fig. 18 shows the evolution of the normalized largest principal stress in the rock during the impact with respect to the normalized initial kinetic energy of the piston. The initial kinetic energy is changed by three approaches separately, i.e., changing the piston length, changing the piston radius and changing the initial velocity of the piston, while keeping the other parameters fixed at the middle values of their domains. Numerical results from the three approaches are depicted by three curves in Fig. 18. All these curves indicate that increasing the initial kinetic energy can increase the largest principal stress in the rock, which is not a surprise. For the purpose of increasing the largest stress in the rock over $800\ MPa$ for $E_p = 200\ GPa$ through increasing the initial kinetic energy, Fig. 18 indicates that the most effective way is to increase the piston's initial velocity.

# 5. Conclusions

The hydraulic impact problem of a non-explosive rock breaking technology has been studied. Dimensional analysis and the finite element method have been applied to systematically investigate the influence of all the parameters involved in the impact process, which includes the geometrical parameters and the properties of rock, piston and water. Major conclusions from our investigation are summarized below:

- The rock density has a negligible influence on the largest principal stress in the rock.

- The influences of rock's Poisson's ratio and Young's modulus on the largest principal stress in the rock are small.

- The shape of the bottom of the hole has a substantial impact on the problem. For example, the closer it is to a right angle, the larger will be the largest stress in the rock.

- The largest principal stress in the rock decreases if the depth of water is increased.

- Increasing the initial kinetic energy of the piston has a significant influence on the problem: it implies an increase of the largest principal stress in the rock and a variation of the time to reach that value.

- The best way to increase the largest principal stress in the rock by increasing the initial kinetic energy of the piston is to increase its initial velocity.

# Acknowledgement

This work has been partially supported by the Australia Research Council. Many thanks to Laurent Champaney for his useful Perl scripts.

**Table 1.** List of all the parameters involved in the impact simulation.

| Piston | Water | Rock | Borehole |
|---|---|---|---|
| $E_p$ : Young's modulus | $E_w$ : Young's modulus | $E_r$ : Young's modulus | $R_h$ : radius |
| $\nu_p$ : Poisson's ratio | $\nu_w$ : Poisson's ratio | $\nu_r$ : Poisson's ratio | $D_h$ : depth |
| $\rho_p$ : density | $\rho_w$ : density | $\rho_r$ : density | $R_a$ : arc radius |
| $L_p$ : length | $D_w$ : depth | | |
| $V_p$ : initial velocity | | | |

**Table 2.** Mechanical properties of typical rock materials.

| Rock material | Density (Kg / m$^3$) | Young modulus (GPa) | Poisson's ratio |
|---|---|---|---|
| Granite | 2500 - 2800 | 35 - 80 | 0.1 - 0.2 |
| Basalt | 2400 - 2900 | 20 - 100 | 0.1 - 0.3 |
| Sandstone | 2200 - 2700 | 10 - 40 | 0.2 - 0.3 |
| Dolerite | 2900 - 3100 | 40 - 90 | 0.1 - 0.3 |
| Limestone | 2000 - 2800 | 10 - 50 | 0.2 - 0.35 |
| Andesine | 2500 - 2800 | 30 - 60 | 0.1 - 0.25 |

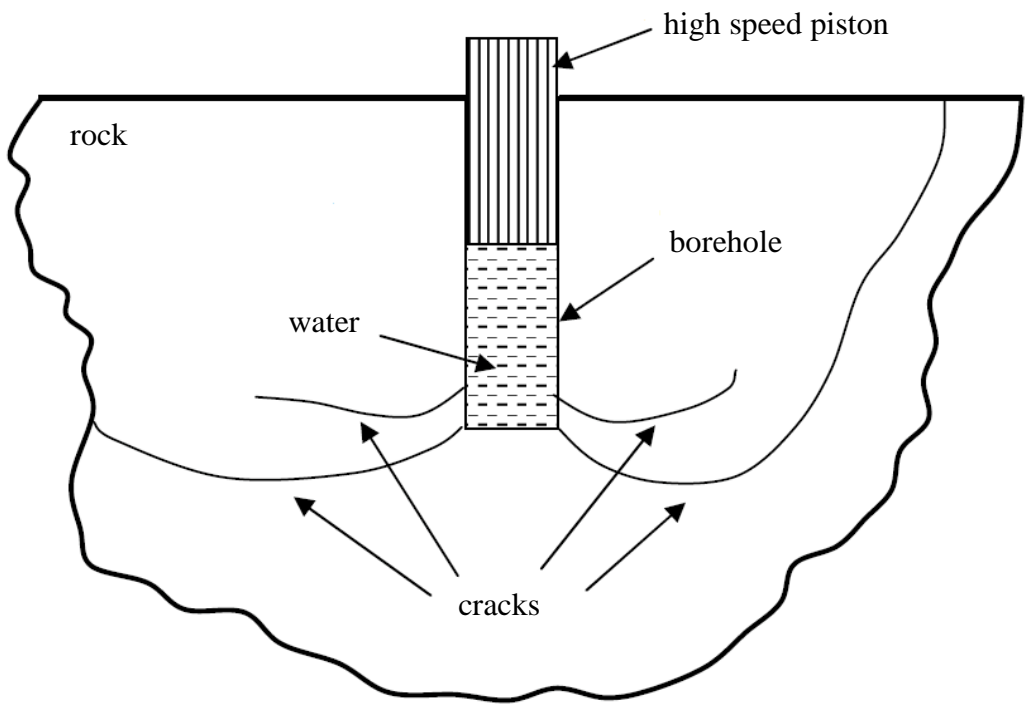

**Fig. 1.** Illustration of the hydraulic rock breaking technology.

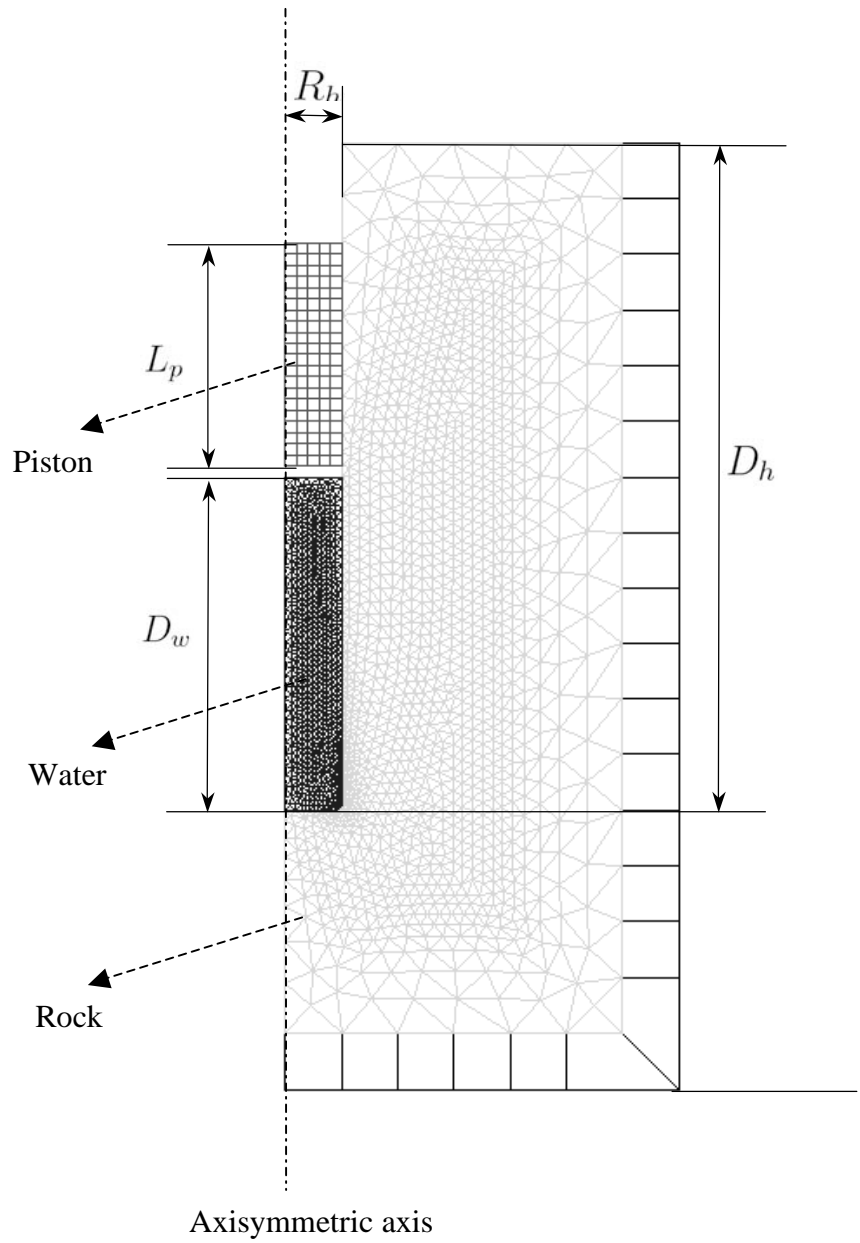

Fig. 2(a)

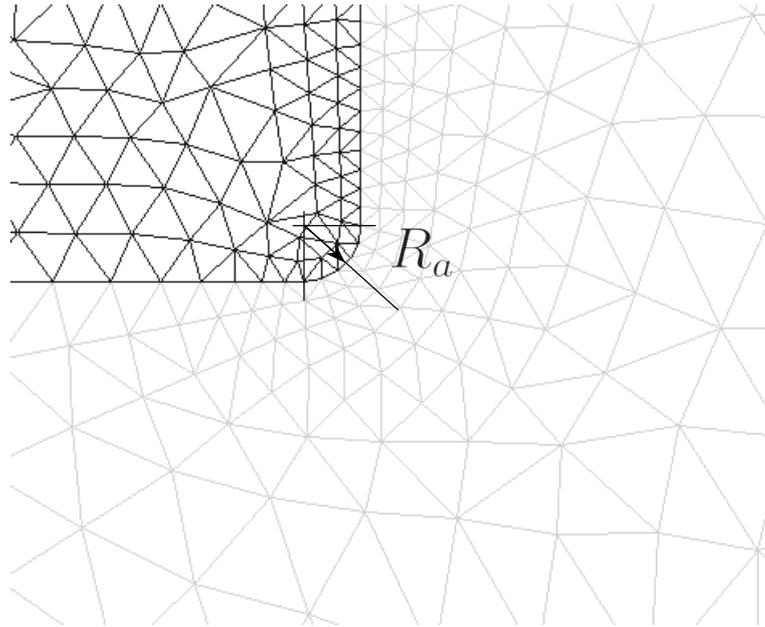

Fig. 2(b)

**Fig. 2.** Finite element mesh and relevant geometrical parameters: (a) global mesh; (b) local mesh around the bottom of the borehole.

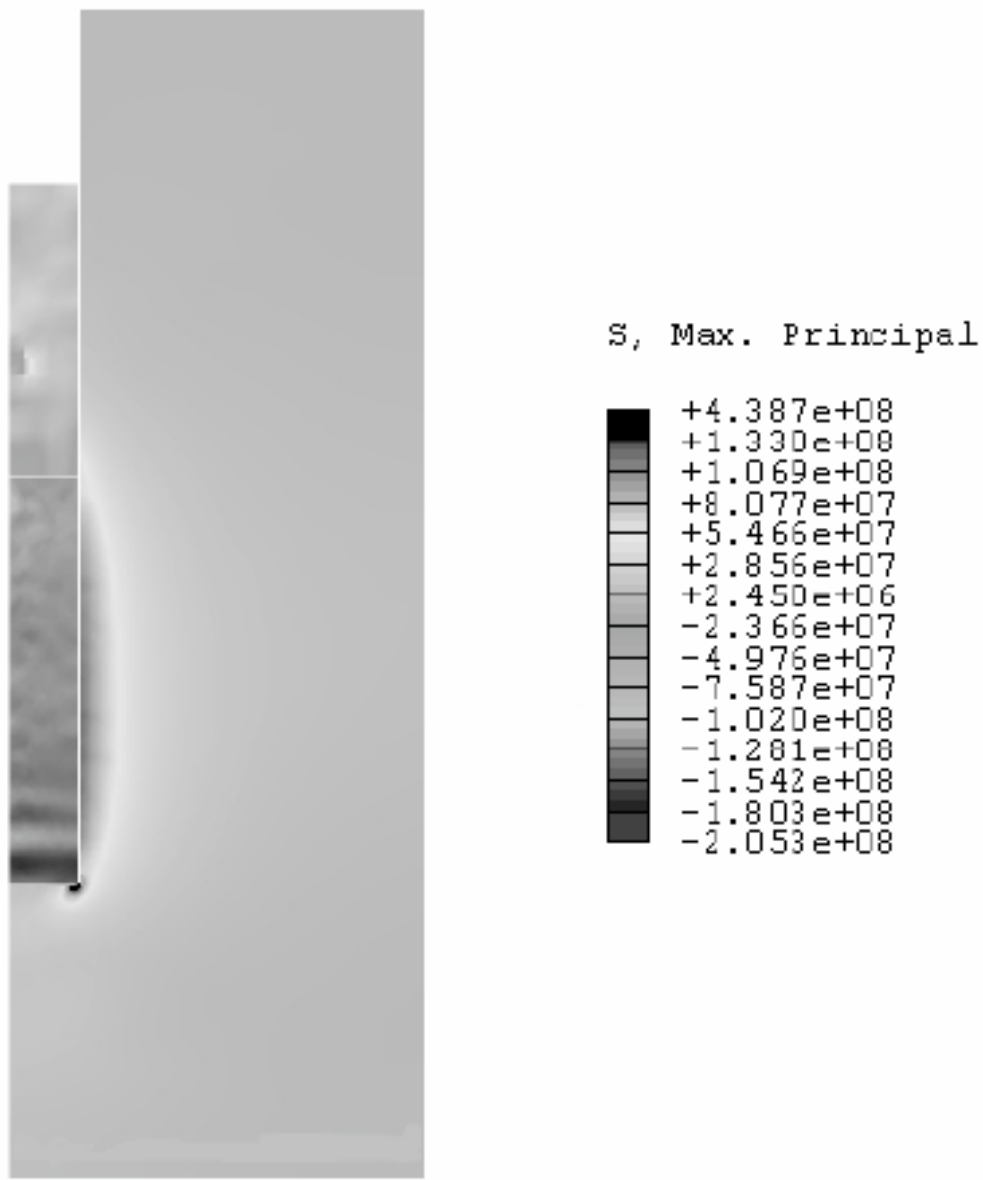

**Fig. 3.** Maximum principal stress field, in the structure.

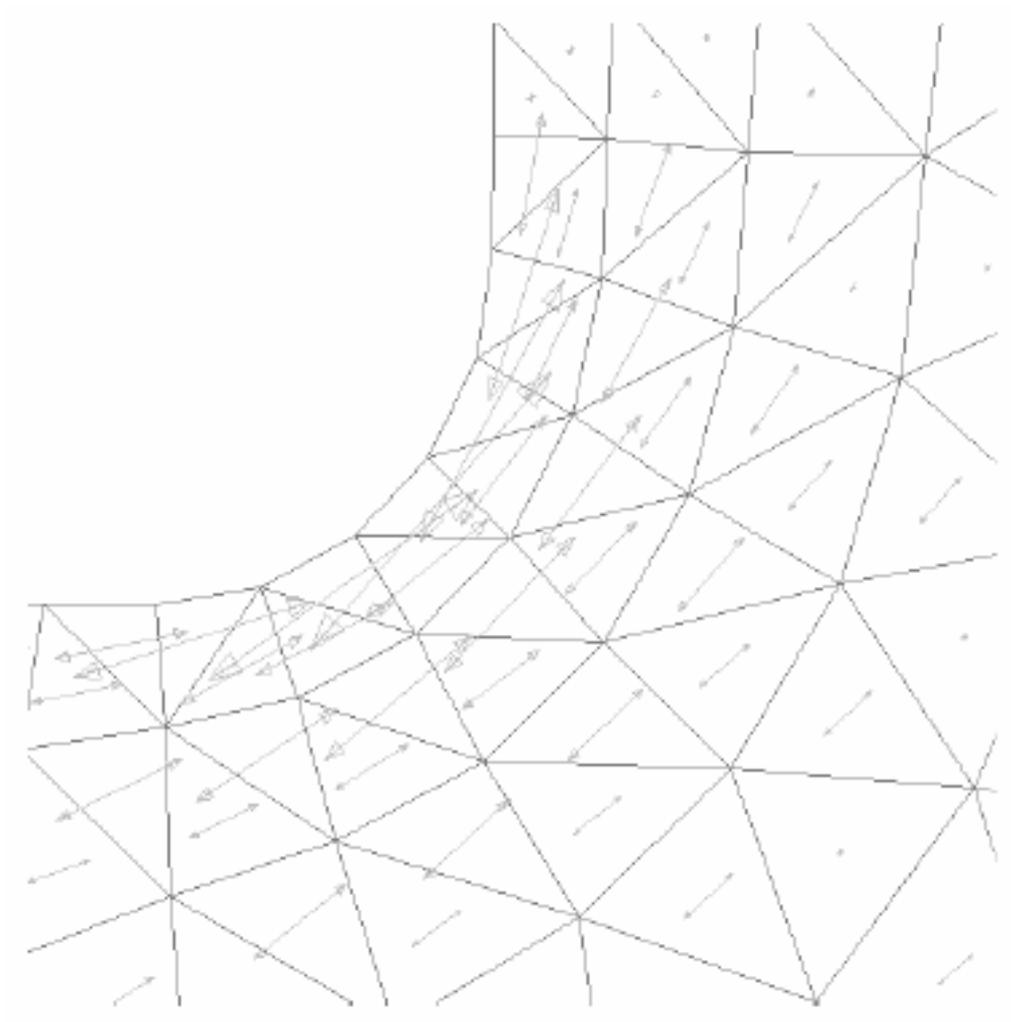

**Fig. 4.** Direction field of the maximum principal stress, in the rock at the bottom of the hole.

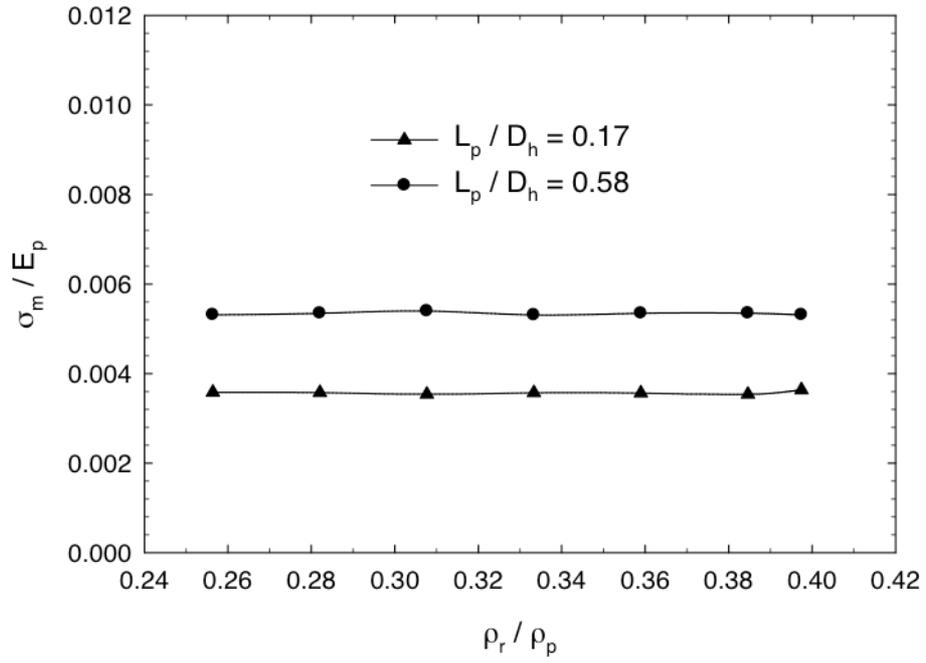

Fig. 5(a)

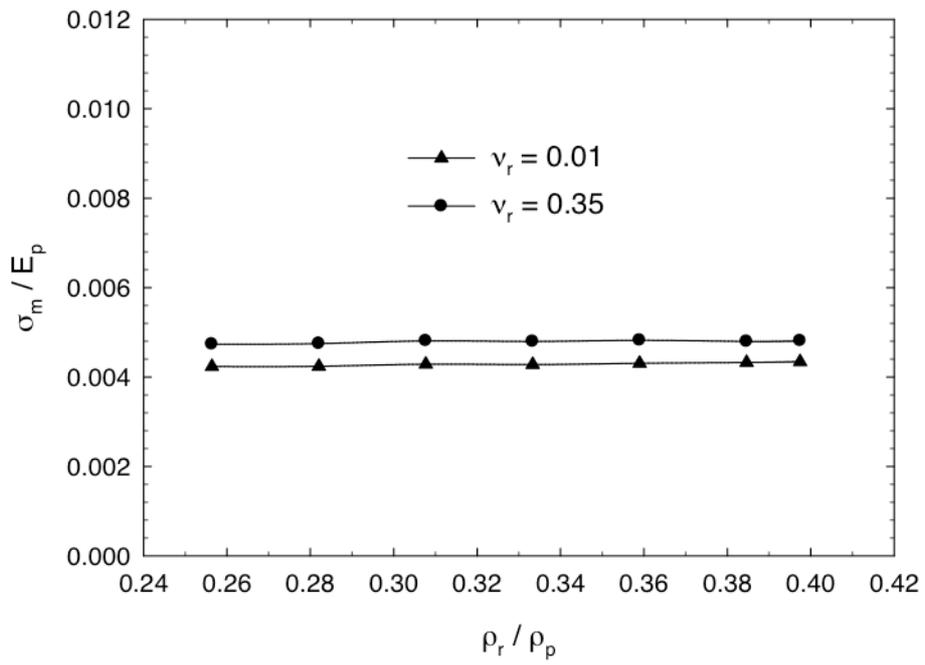

Fig. 5(b)

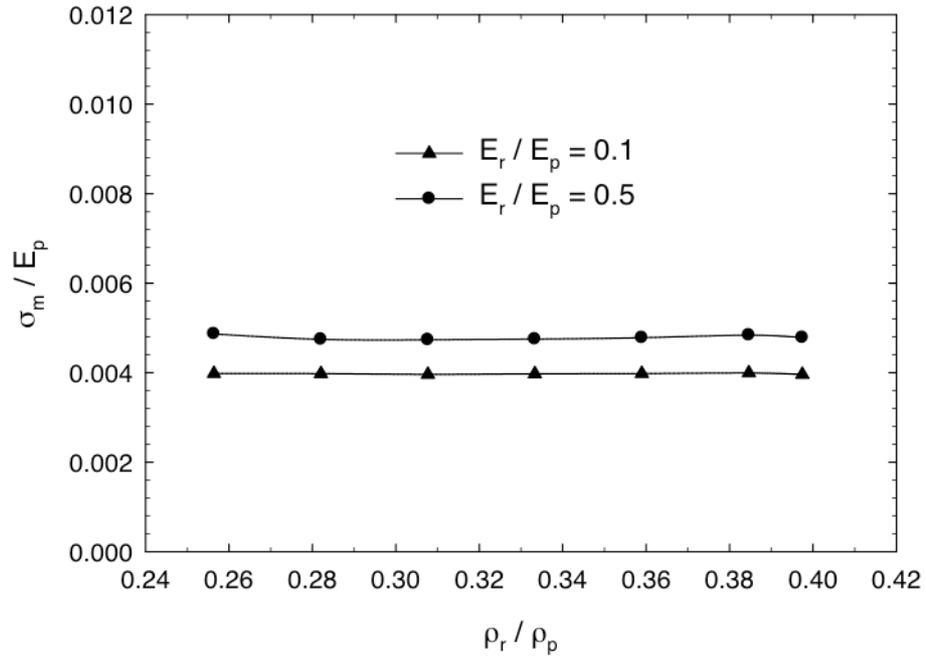

Fig. 5(c)

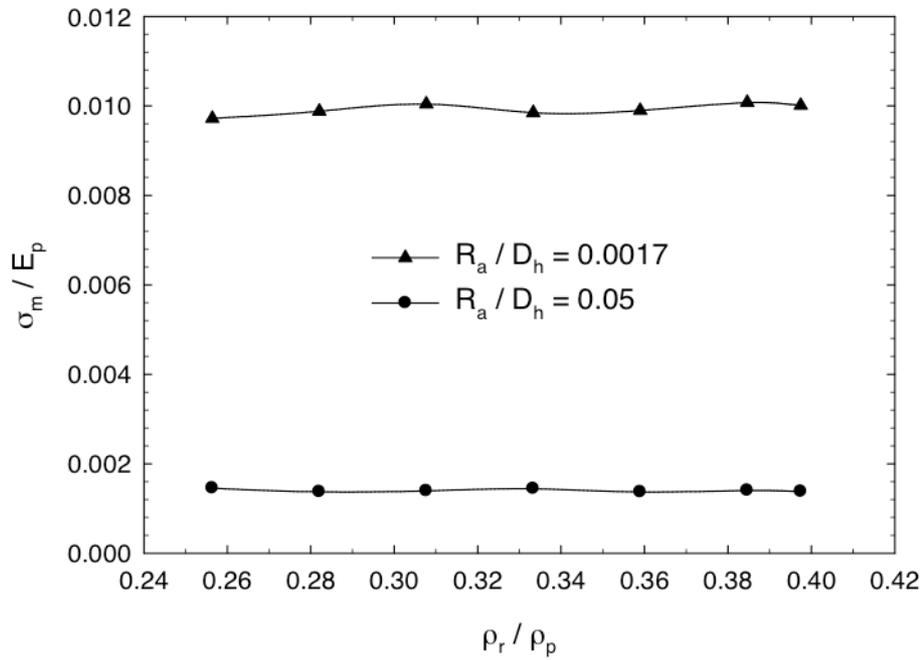

Fig. 5(d)

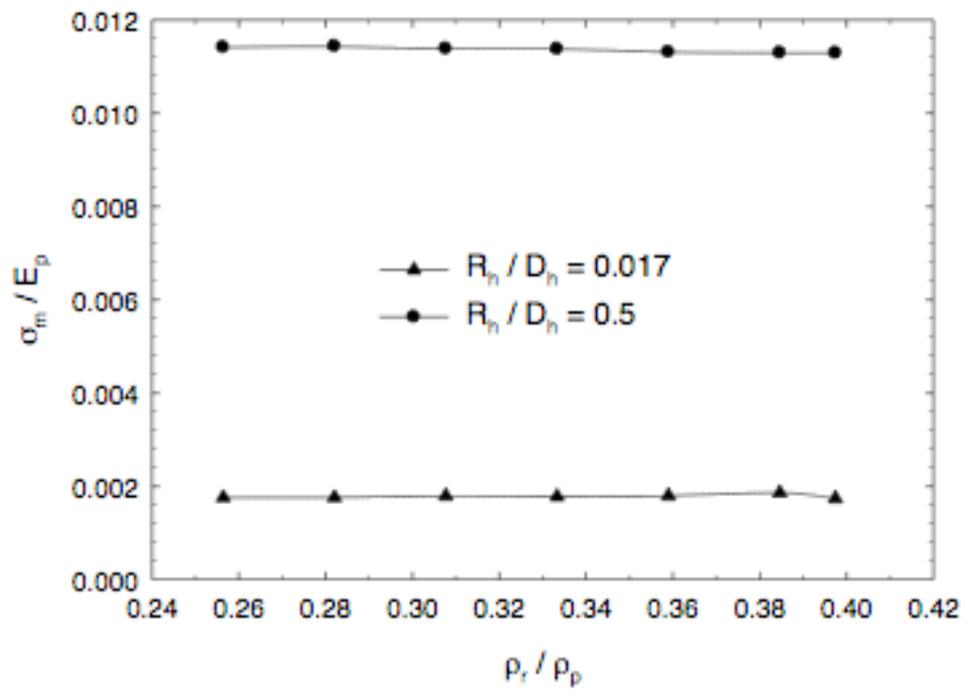

Fig. 5(e)

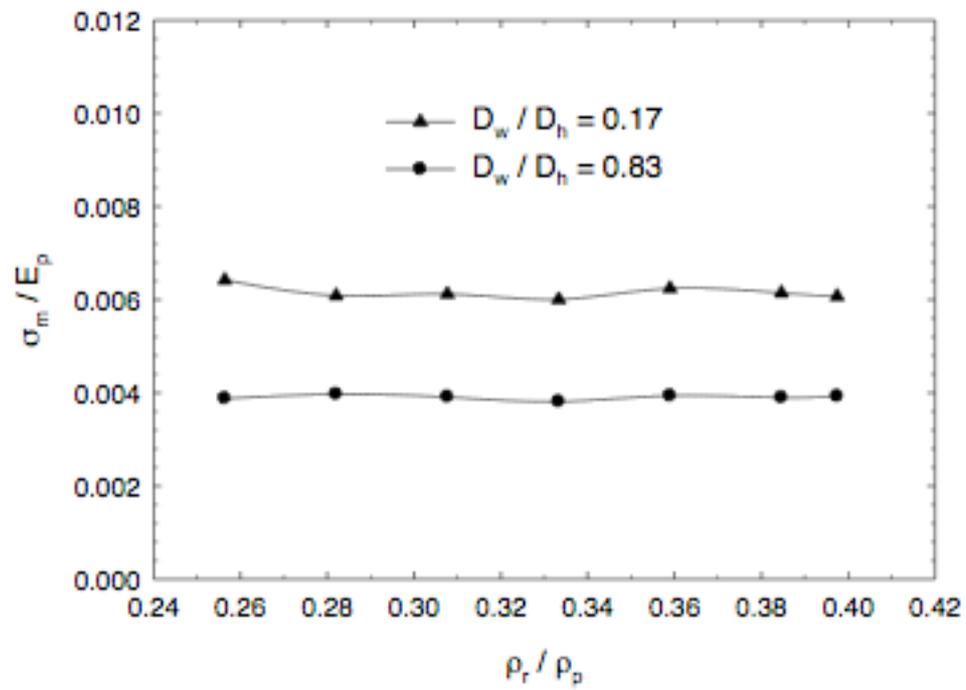

Fig. 5(f)

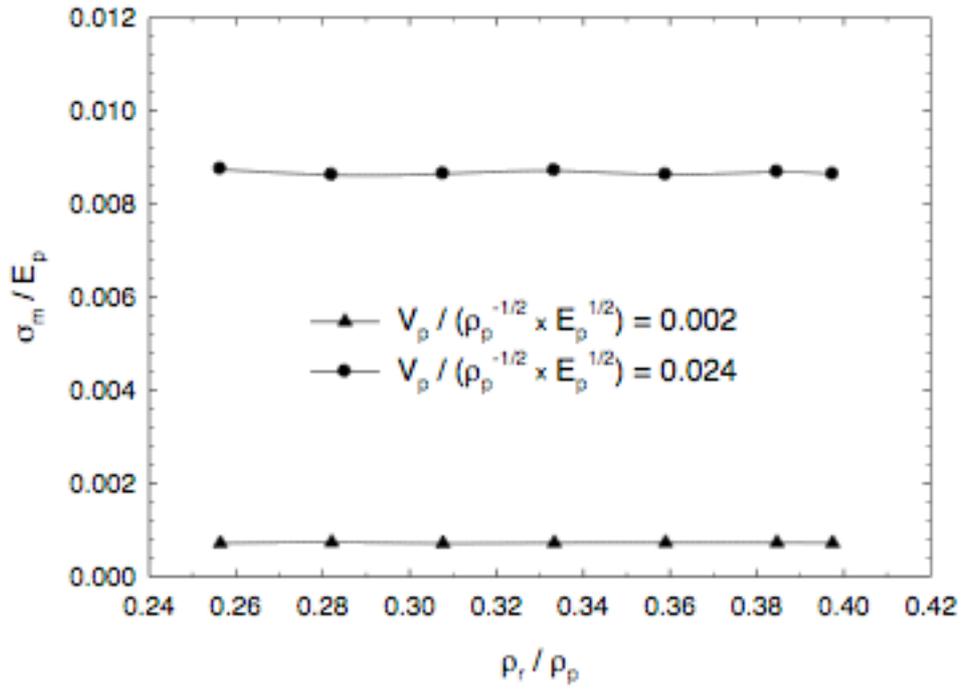

Fig. 5(g)

**Fig. 5.** Influence of normalized rock density on the normalized largest principal stress in the rock for the cases (a) $L_p/D_h = 0.17$ and $L_p/D_h = 0.58$; (b) $v_r = 0.01$ and $v_r = 0.35$; (c) $E_r/E_p = 0.1$ and $E_r/E_p = 0.5$; (d) $R_a/D_h = 0.0017$ and $R_a/D_h = 0.05$; (e) $R_h/D_h = 0.017$ and $R_h/D_h = 0.5$; (f) $D_w/D_h = 0.17$ and $D_w/D_h = 0.83$; (g) $V_p/(\rho_p^{-1/2}E_p^{1/2}) = 0.002$ and $V_p/(\rho_p^{-1/2}E_p^{1/2}) = 0.024$.

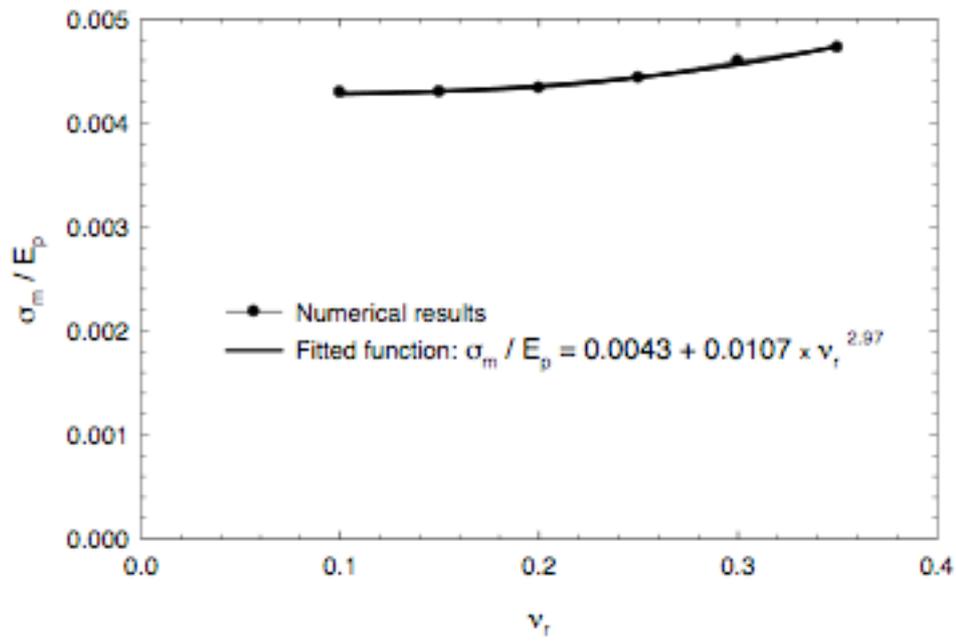

**Fig. 6.** Influence of Poisson's ratio of the rock on the normalized largest principal stress in the rock during the impact.

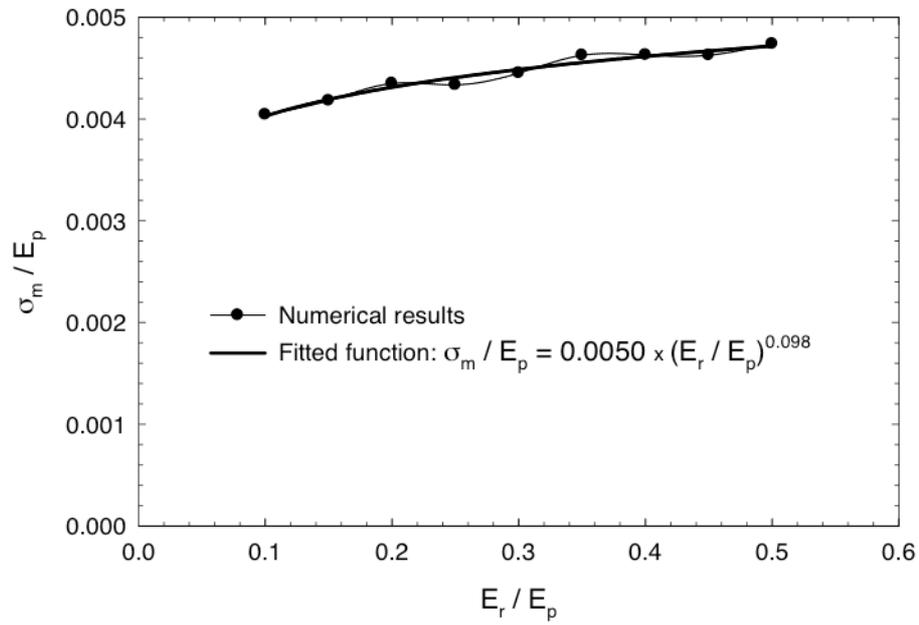

**Fig. 7.** Influence of normalized Young's modulus of the rock on the normalized largest principal stress in the rock during the impact.

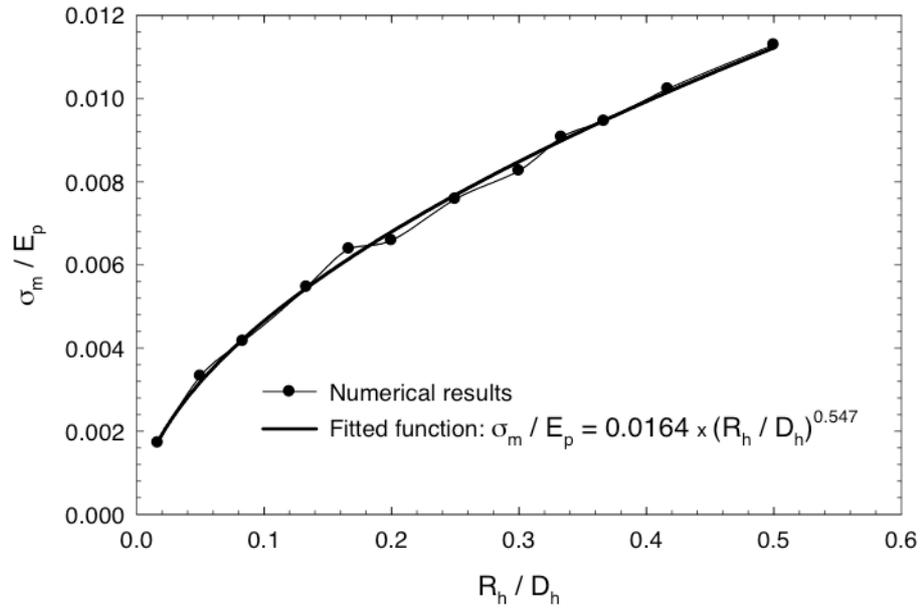

**Fig. 8.** Influence of normalized borehole radius on the normalized largest principal stress in the rock during the impact.

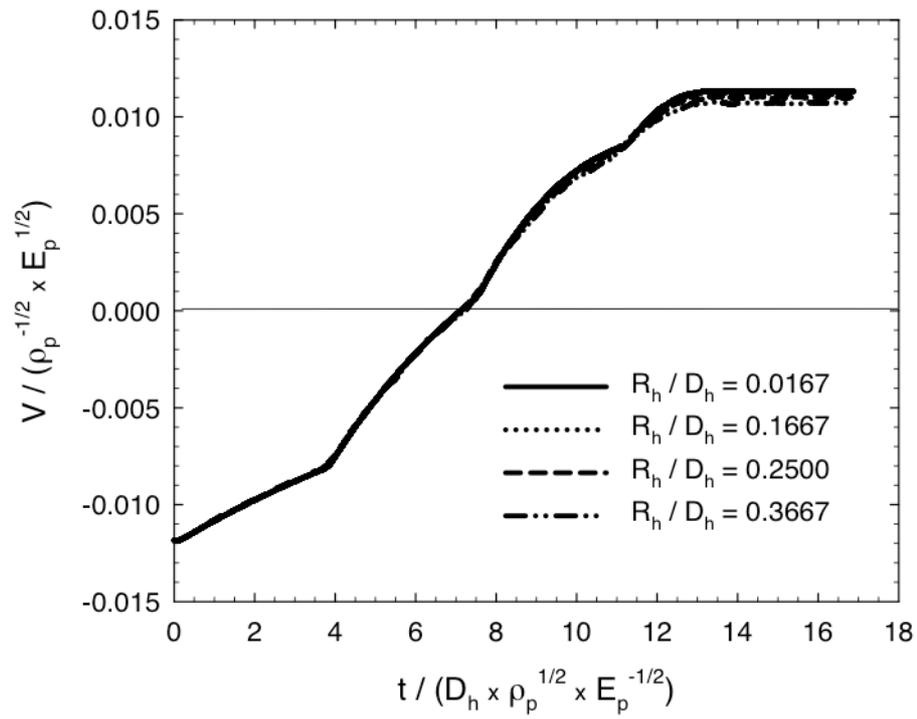

**Fig. 9.** Normalized average velocity of the piston versus normalized impacting time for several values of normalized borehole radius.

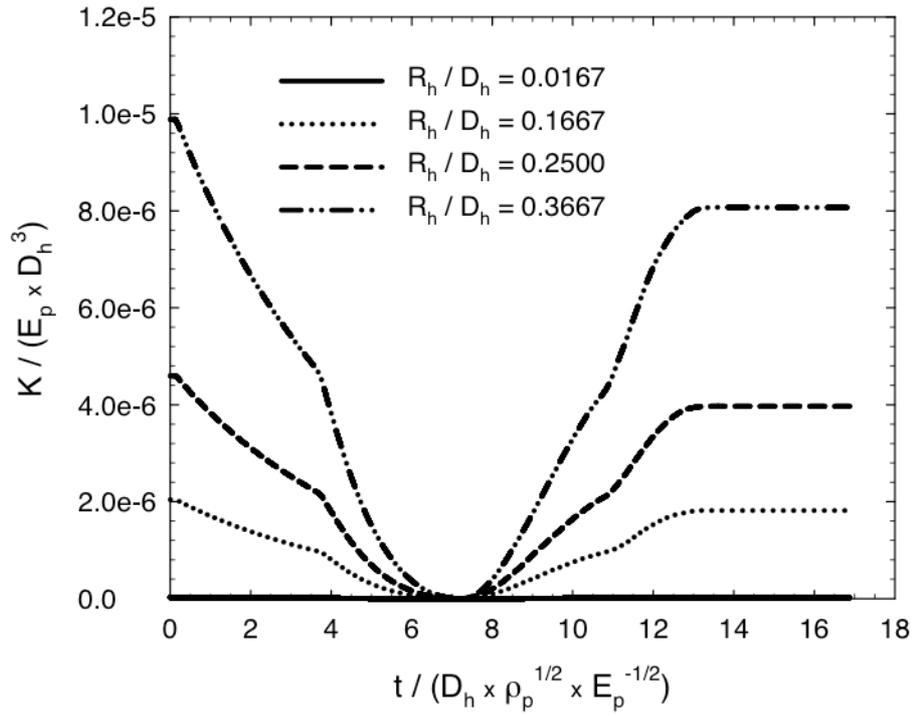

Fig. 10. Normalized kinetic energy of the piston versus normalized impacting time for several values of normalized borehole radius.

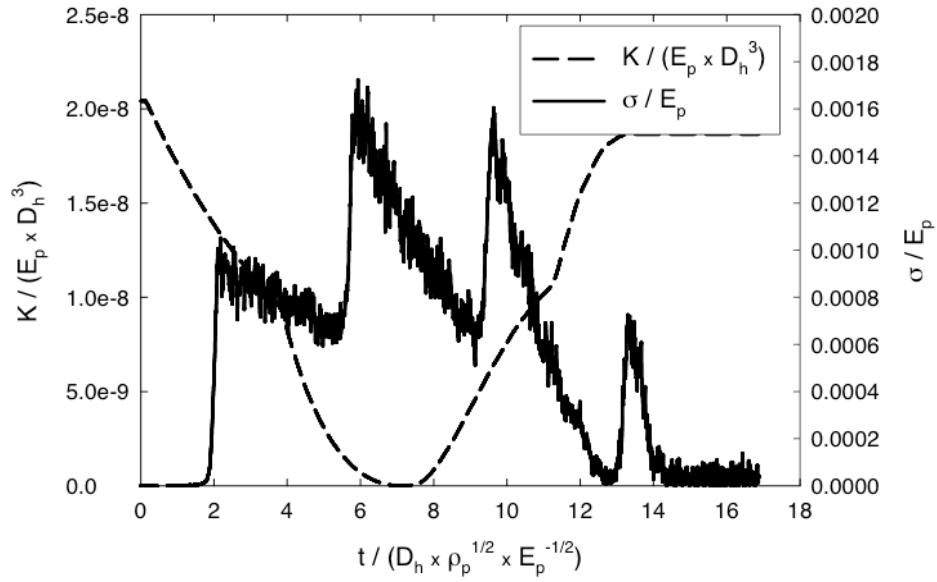

**Fig. 11.** Normalized energy of the piston and the normalized maximum principal stress in the element that has the largest principal stress in the rock during the impact versus normalized impacting time, for $R_h/D_h = 0.0167$.

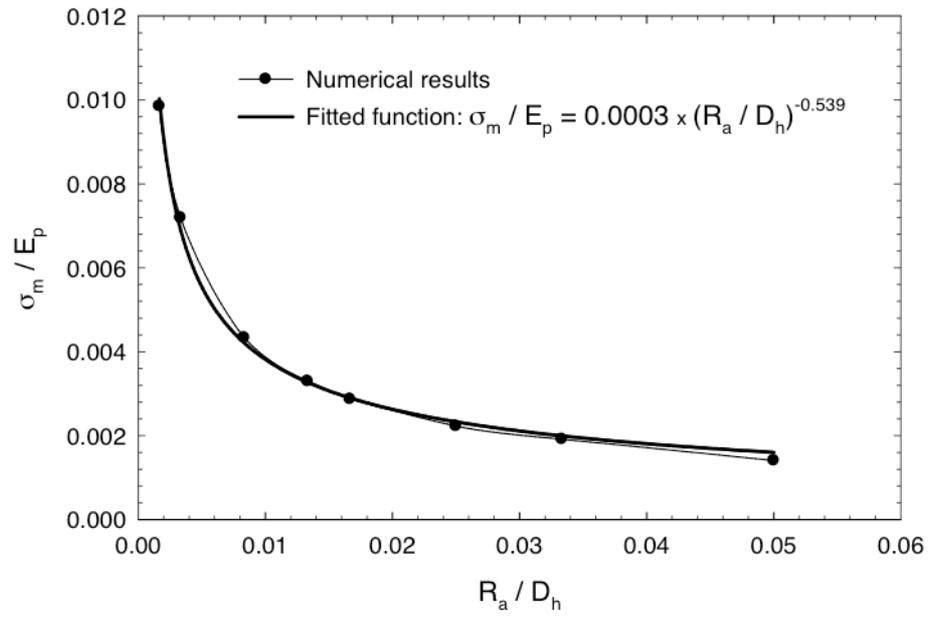

**Fig. 12.** Influence of normalized arc radius at the bottom of the hole on the normalized largest principal stress in the rock during the impact.

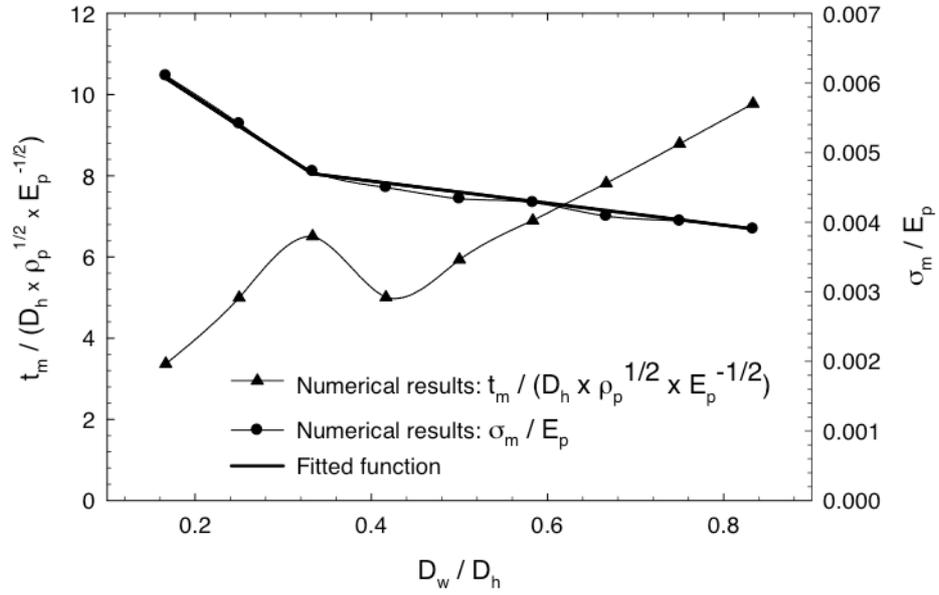

**Fig. 13.** Influence of normalized water depth on the normalized largest principal stress in the rock during the impact and normalized impacting time to reach that stress.

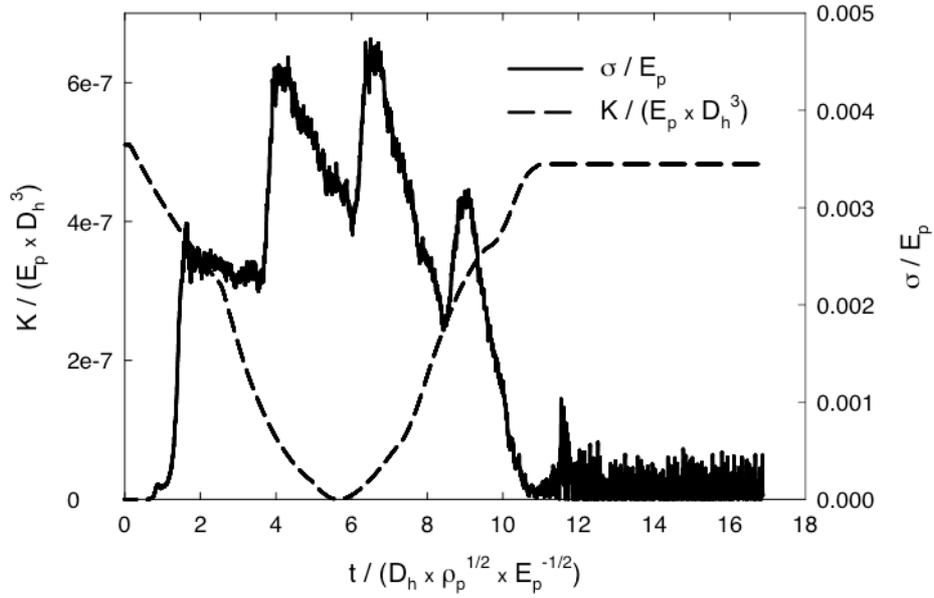

Fig. 14(a)

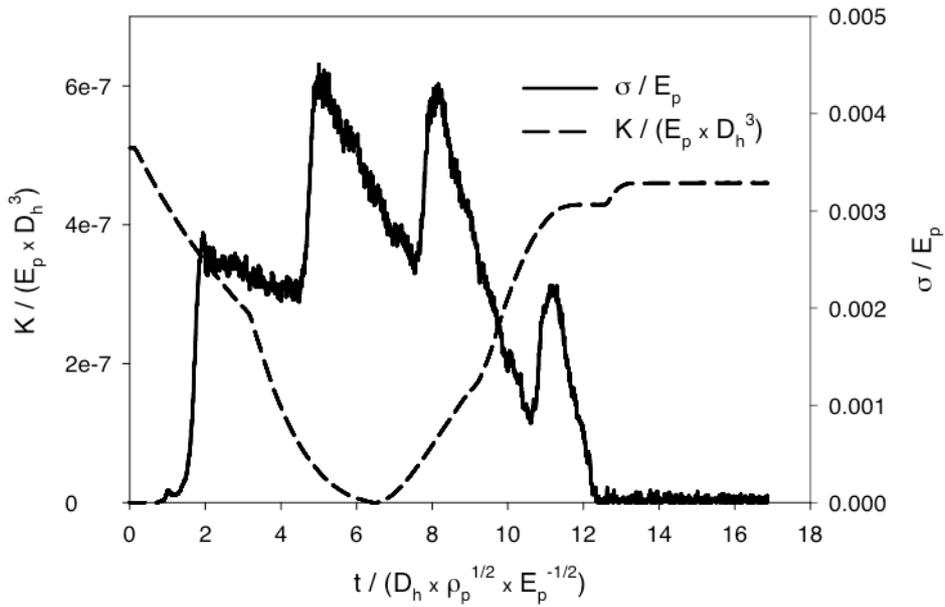

Fig. 14(b)

**Fig. 14.** Normalized kinetic energy of the piston and the normalized maximum principal stress in the element that has the largest principal stress in the rock during the impact versus normalized time for (a) $D_w/D_h = 0.333$ and (b) $D_w/D_h = 0.417$.

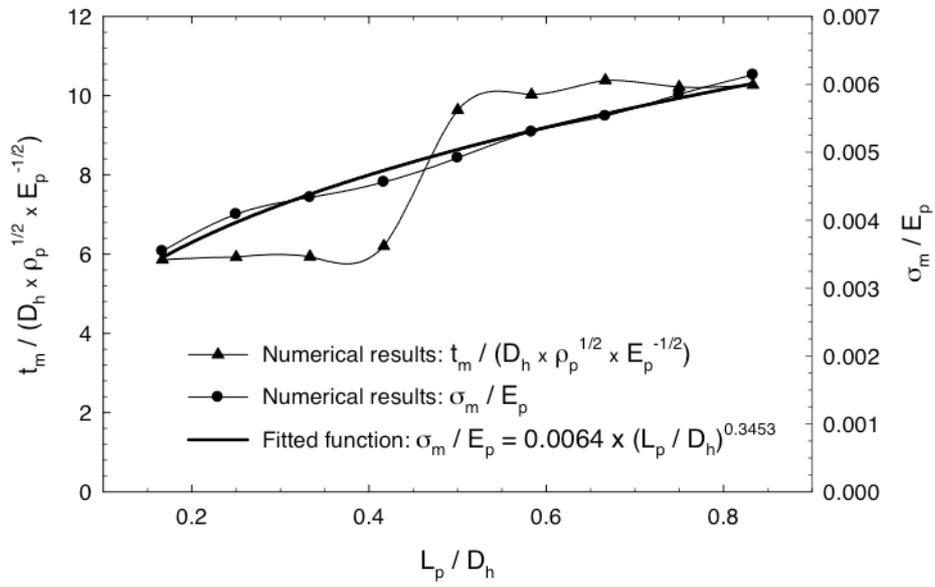

**Fig. 15.** Influence of normalized piston length on the normalized largest principal stress in the rock and the normalized time to reach that stress.

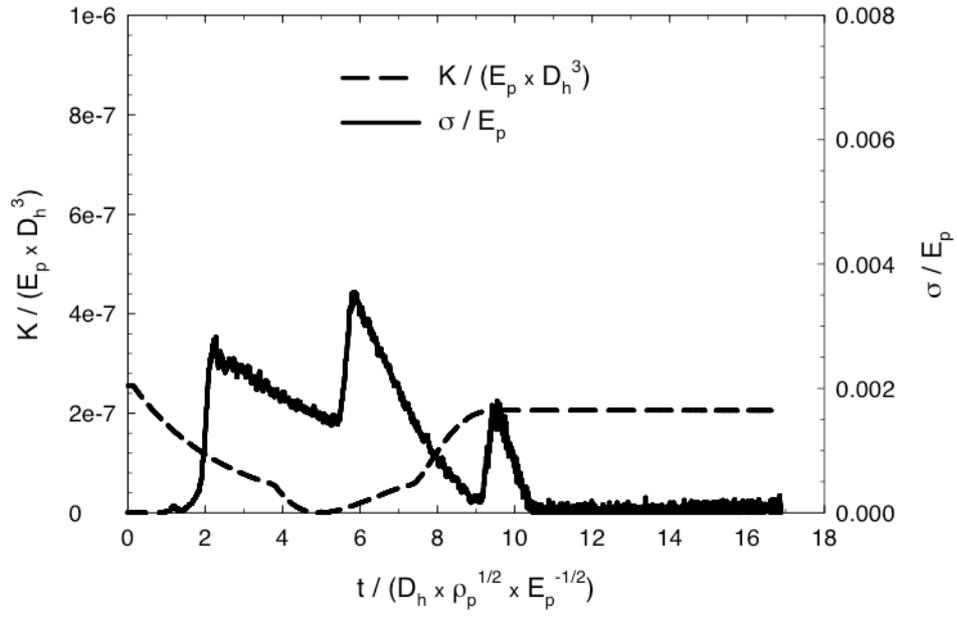

Fig. 16(a)

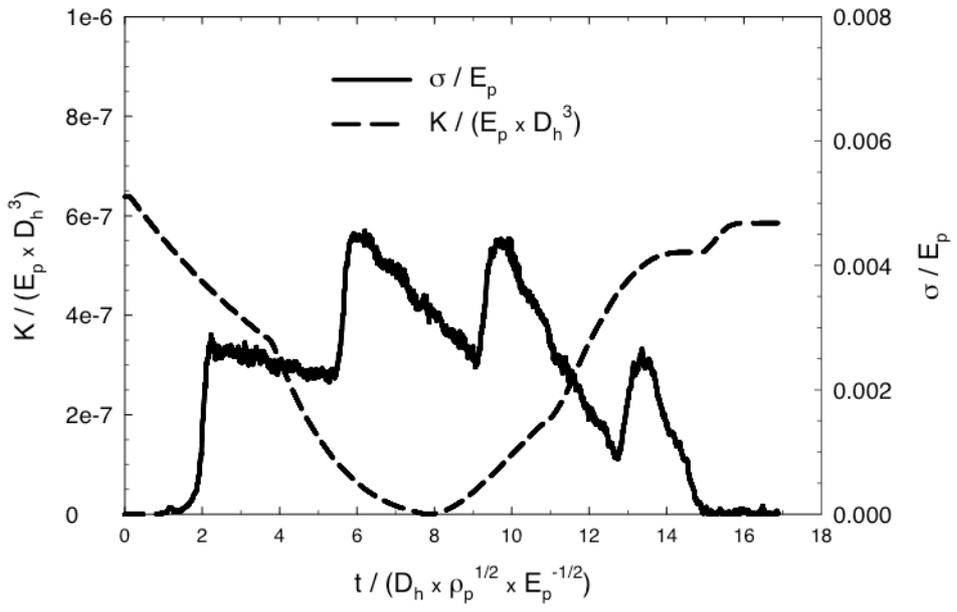

Fig. 16(b)

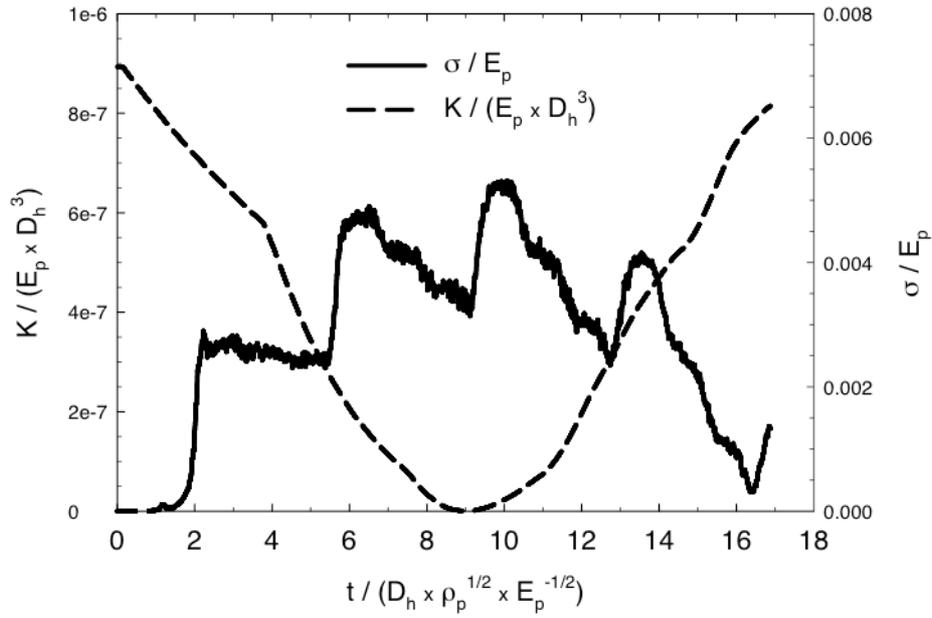

Fig. 16(c)

**Fig. 16.** Normalized kinetic energy of the piston and normalized first principal stress in the element that has the largest principal stress in the rock during the impact versus normalized impacting time for (a) $L_p/D_h = 0.1667$, (b) $L_p/D_h = 0.4167$ and (c) $L_p/D_h = 0.5833$.

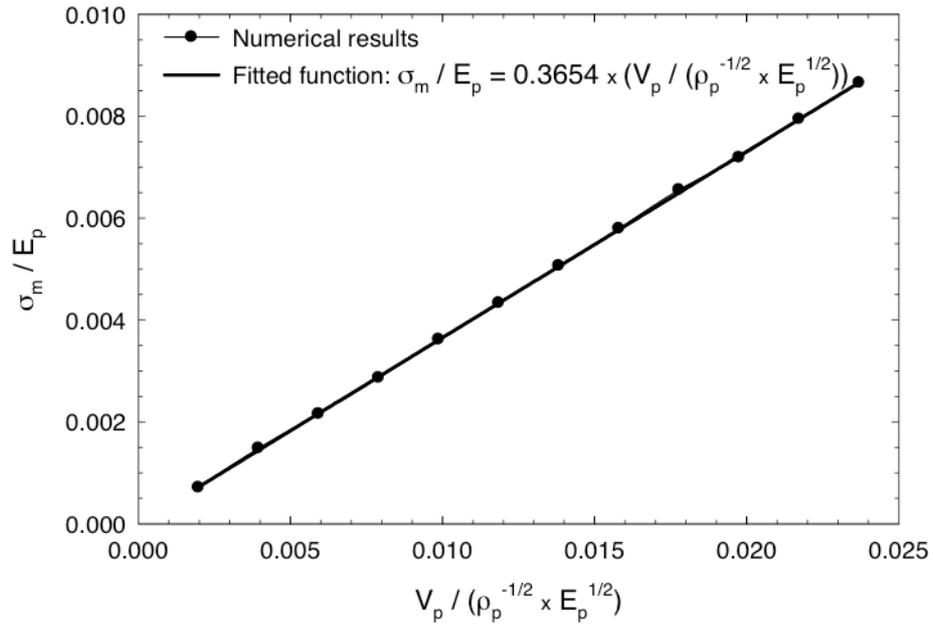

**Fig. 17.** Influence of normalized initial velocity of the piston on the normalized largest principal stress in the rock during the impact.

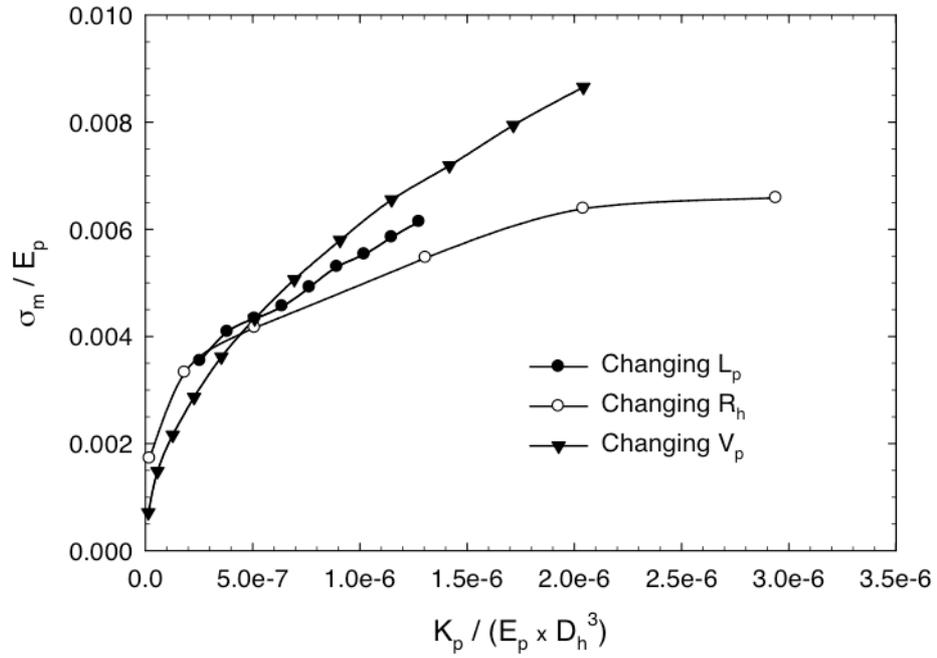

**Fig. 18.** Influence of normalized initial kinetic energy of the piston on normalized largest principal stress in the rock during the impact applying different ways to increase the initial kinetic energy of the piston.